\documentclass[twocolumn]{aastex63}
\usepackage{amsmath,amstext,amssymb}

\newcommand{\Milwaukee}{University of Wisconsin-Milwaukee, Milwaukee, WI 53201, USA}

\newcommand{\T}{{\mbox{\tiny T}}}

\newcommand{\V}{{\mbox{\tiny V}}}
\newcommand{\R}{{\mbox{\tiny R}}}

\shorttitle{Spacetime Dimensions from GWTC-3}
\shortauthors{Maga\~na~Hernandez}

\graphicspath{{./}{figures/}}

\begin{document}

\title{Constraining the number of spacetime dimensions from GWTC-3 binary black hole mergers}
\correspondingauthor{Ignacio Maga\~na~Hernandez}
\email{maganah2@uwm.edu}
\author[0000-0003-2362-0459]{Ignacio Maga\~na~Hernandez}
\affiliation{\Milwaukee}

\begin{abstract}
In modified gravity models that allow for additional non-compact spacetime dimensions, energy from gravitational waves can leak into these extra spacetime dimensions, leading to a reduction in the amplitude of the observed gravitational waves, and thus a source of potential systematics in the inferred luminosity distances to gravitational wave sources. Since binary black hole (BBH) mergers are standard sirens, we use the pair-instability supernova (PISNe) mass gap and its predicted features to determine a mass scale and thus be able to break the mass-redshift degeneracy. We simultaneously fit for the BBH population and the extra spacetime dimensions parameters from gravitational leakage models using BBH observations from the recently released GWTC-3 catalog. We set constraints on the number of spacetime dimensions and find that $D= 3.95^{+0.09}_{-0.07}$ at $68\%$ C.L. for models that are independent of a screening scale, finding that the GWTC-3 constraint is as competitive as that set from GW170817 and its electromagnetic counterpart. For models where gravity leaks below a certain screening scale $R_c$, we find $D=4.23^{+1.50}_{-0.57}$ and $\log_{10} R_c/{\rm Mpc}= 4.14^{+0.55}_{-0.86}$ with a transition steepness $\log_{10} n = 0.86^{+0.73}_{-0.84}$ for the leakage, which for the first are constrained jointly with the BBH population at cosmological distances. These constraints are consistent with General Relativity (GR) where gravitational waves propagate in $D=3+1$ spacetime dimensions. Using the BBH population to probe modifications to standard cosmological models provides an independent test of GR that does not rely on any electromagnetic information but purely on gravitational wave observations.
\end{abstract}


\section{Introduction}
With the recent release of the GWTC-3 catalog \citep{gwtc1,gwtc2,gwtc21,gwtc3} from the LIGO Scientific, Virgo and KAGRA Collaborations(LVK), the number of gravitational wave detections from binary black hole mergers has increased to 90 confidently detected events \citep{LIGOScientific:2014pky,VIRGO:2014yos}. The increasing size of GW catalogs has enabled the study of the BBH population \citep{o3a_pop,o3pop}, its cosmic expansion history \citep{o2cosmo,o3cosmo}, signatures of gravitational wave lensing \citep{o3lens} and how well the population agrees with General Relativity \citep{o3atgr,LIGOScientific:2021sio}. 

Gravitational wave sources provide a direct measurement of their luminosity distance---they are standard sirens \citep{Schutz:1986gp, Holz:2005df}. Uniquely associated electromagnetic counterparts can constrain the redshift of the source and hence allow for an independent determination of the corresponding electromagnetic luminosity distance, as was the case for the first bright standard siren GW170817 \citep{TheLIGOScientific:2017qsa}. Combined GW and EM counterpart measurements have allowed constraints (to mention a few) of cosmological parameters such as the Hubble constant \citep{Abbott:2017xzu}, the number of extra spacetime dimensions allowed under gravitational leakage models \citep{Pardo:2018ipy} and modified gravitational wave propagation due to a running Planck mass \citep{PhysRevD.99.083504}. 

For gravitational wave events without expected EM counterparts (dark sirens), including the numerous binary black hole mergers, in absence of a uniquely identified host galaxy, a galaxy survey can be used as prior information on the potential host galaxies of the event in combination with its gravitational wave localization volume~\citep{DelPozzo:2011yh,Nair:2018ign,Chen:2017rfc,Fishbach:2018gjp,Gray:2019ksv,Soares-Santos:2019irc,Abbott:2019yzh,DES:2020nay,Mukherjee:2020hyn,Diaz:2021pem}.

The dark siren methodology has been used to constrain  cosmology \citep{Abbott:2017xzu,Soares-Santos:2019irc,DES:2020nay,o3cosmo,Palmese:2021mjm} as well as to measure modified gravitational wave propagation \citep{Finke:2021aom,Mukherjee:2020mha}.

However, even without electromagnetic information, one can still make a statistical measurement of redshift using the features of the population distribution of compact binary mergers. Since we measure redshifted ``detector-frame" masses, $m_1^{\rm det} = m_1(1+z)$, one can model the expected source frame mass distribution (given our current understanding on BBH formation channels) on $m_1$ to estimate the redshift $z$. This idea was first explored in \cite{PhysRevD.85.023535} in the context of binary neutron star mergers and 3G detectors to constrain cosmology, where the ``known" galactic neutron star mass distribution (a sharply peaked Gaussian with mean around 1.4 $M_{\odot}$) was used as a feature to break the mass-redshift degeneracy.

More recently, \cite{Farr_2019}, applied this methodology to BBH mergers by using the predicted cut off and excess of black holes with masses $M\approx 45 M_{\odot}$ in the BH mass spectrum due to the pair-instability supernovae (PISNe) mass gap \citep{PISN_Woosley,Heger_2002,Heger_2003,Woosley_2017,Woosley_2019}. Recent work in \cite{Ezquiaga:2021ayr}, used the same methodology to constrain the value of $c_M$ (in general relativity, $c_M = 0)$, a parameter that allows for modifications of $\Lambda$CDM due to a time-varying Planck mass \citep{PhysRevD.99.083504} as well as other modified gravity parameterizations \citep{Mancarella:2021ecn}. The LVK collaboration, subsequently applied the methodology of \cite{Farr_2019} to the GWTC-3 catalog to constraint the cosmic expansion history of BBH mergers using astrophysically motivated source frame mass models \citep{o3cosmo}.

In this work, we set constraints on gravitational leakage models, specifically, on the number of extra non-compact spacetime dimensions where gravity may leak, using the observed population of BBH mergers in GWTC-3. In order to determine the redshift for these events, we fit the BBH population with an astrophysically motivated PISNe mass model together with modifications to the gravitational wave luminosity distance induced by the gravitational leakage models we consider.  

This paper is organized as follows. In Section 2 we describe the gravitational leakage models and how these relate to the damping of the gravitational wave amplitude. In Section 3, we summarize the hierarchical Bayesian framework used in our analysis and provide the details for the BBH population model that we use. In Section 4, we present the main results of this paper and in Section 5, we provide a summary of this work. We use the Planck 2015 cosmological model throughout this paper, that is, $H_0 = 67.8 \ \rm km/s/Mpc$, $\Omega_{m,0} = 0.308$ and set $\Omega_{k,0} = 0$.

\section{Gravitational Leakage Models}
\label{sec:leak}
In this section we describe the different gravitational leakage models and how these relate to higher spacetime dimensional theories. We consider how these models modify the gravitational waveform amplitude, and hence correspondingly can bias the observed luminosity distance to the source. This section relies heavily on the work of ~\cite{Deffayet:2007kf,Pardo:2018ipy,Corman:2021avn}.

In General Relativity the gravitational wave strain is proportional to the luminosity distance $d_L^{\rm GW}$ to the source as,
\begin{equation}\label{eqn:grwaveform}
h_{\rm{GR}} \propto \frac{1}{d_L^{\rm{GW}}},
\end{equation}

For a higher-dimensional spacetime theory where there is some leakage of gravity, one would expect damping of the gravitational waveform in the form of a power-law due to flux conservation, so the simplest phenomenological model to consider is \citep{Pardo:2018ipy, Corman:2021avn}: 
\begin{equation}\label{eqn:grwaveform}
d_L^{\rm{GW}} = d_L\left(\frac{d_L}{\textrm{1 Mpc}}\right)^{(D-4)/2},
\end{equation}
where $D$ is the number of spacetime dimensions, $d_L$ is the $D$-dimensional luminosity distance and $d_L^{\rm GW}$ is the measured GW luminosity distance (from parameter estimation analyses). 

However, one usually parameterizes this model so that below a certain length scale the spacetime becomes 4-dimensional \citep{Corman:2021avn}. So following the standard phenomenology, we define a screening scale $R_c$ for the gravitational leakage as well as the overall transition steepness $n$, which determines the strength of the leakage. So more generally, following \citep{Pardo:2018ipy,Corman:2021avn}:

\begin{equation}\label{eqn:defmenwaveform}
d_L^{\rm{GW}} = d_L\left[ 1 + \left(\frac{d_L}{R_c}\right)^{n} \right]^{(D-4)/(2n)},
\end{equation} 
this relation reduces to model in Equation \ref{eqn:grwaveform} for $d_L \gg R_c$ and to $d_L^{\rm{GW}} = d_L$ for $d_L \ll R_c$.

\begin{figure}[htb]
\includegraphics[width=0.5\textwidth]{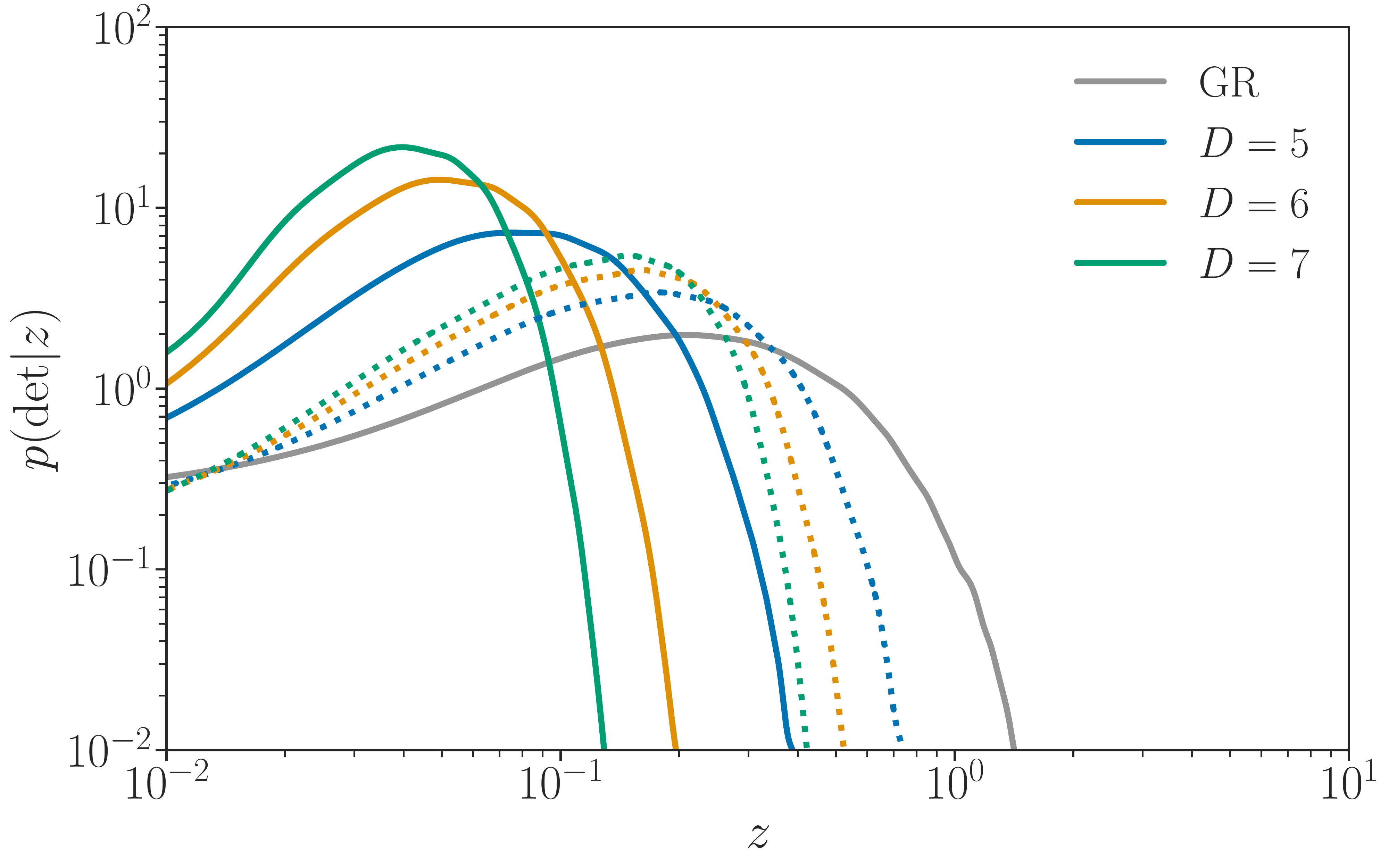}
\includegraphics[width=0.5\textwidth]{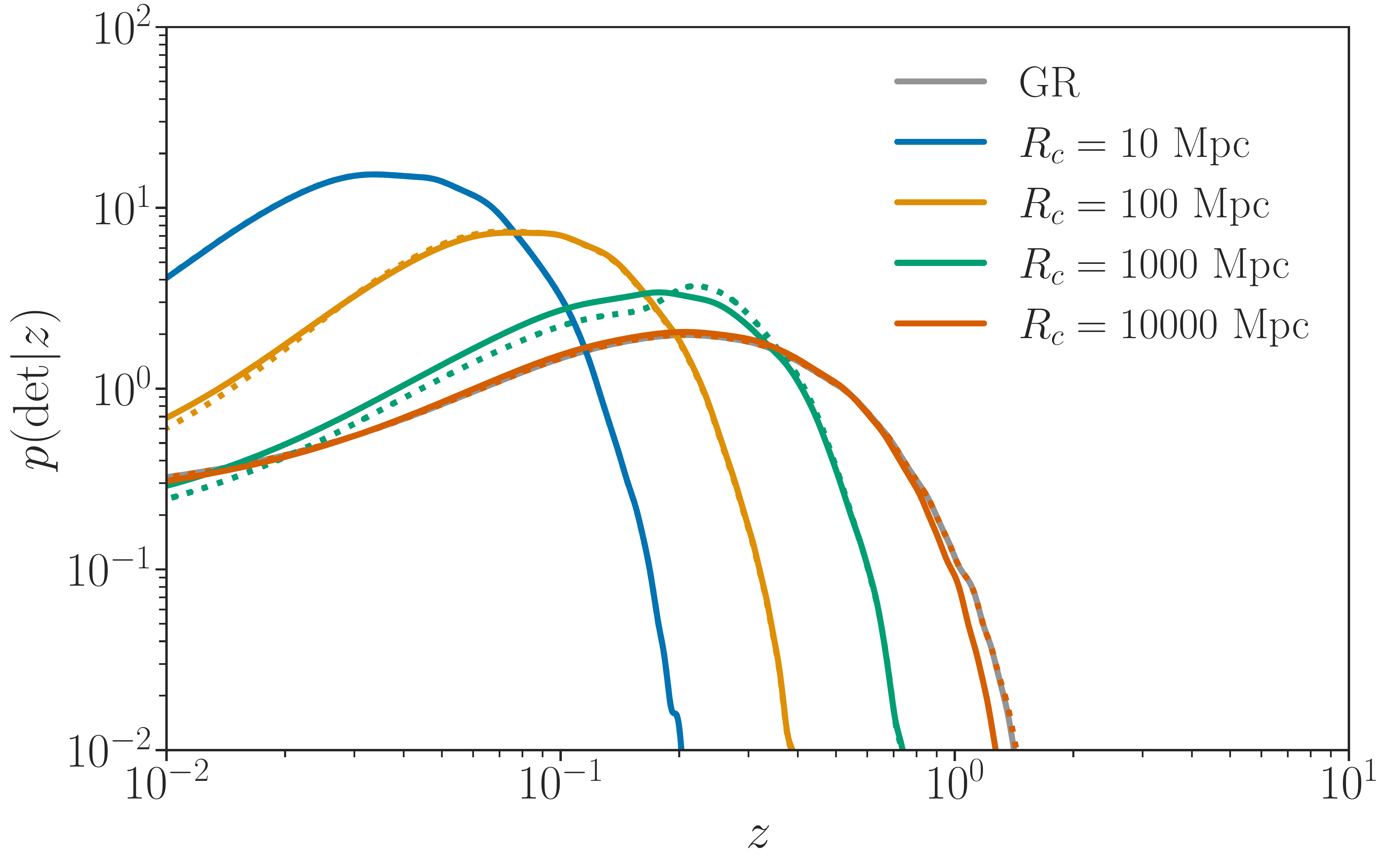}
\caption{Top panel: Detectability of BBH mergers as a function of redshift $z$ when varying the spacetime dimension $D$ at fixed $R_c$ and with $n=1$. Solid lines have a screening scale $R_c=10^2 \rm \ Mpc$, while dashed lines have $R_c=10^3 \rm \ Mpc$ for comparison. Bottom panel: Detectability as a function of redshift given the screening scale $R_c$ at fixed spacetime dimension $D=5$ and transition steepness $n=2$ (solid-lines) and $n=500$ (dashed-lines). We fix the population of BBH mergers to the ``{\textsc{Power Law + Peak}}'' model with parameters $m_{\rm min}=5M_\odot$, $m_{\rm max}=87M_\odot$, $\alpha=1.9$, $\beta=3.4$, $m_{\rm pp} = 35M_\odot$, $\sigma_{\rm pp} = 2M_\odot$ and $f_{\rm pp} = 0.03$ and to the SFR-like redshift evolution model with parameters $\gamma=2.7$, $\kappa=-2$ and $z_{\rm peak}=2$.}
\label{Fig:pdetz_dims}
\end{figure}

To understand how the leakage model of Equation \ref{eqn:defmenwaveform} affects the detectability of a population of BBH mergers, we compute $p(\textrm{det}|z)$, the probability of detecting a BBH at a given redshift $z$, since the corresponding luminosity distance $d_L^{\rm{GW}}$ will change as a function of $D$, $R_c$ and $n$. By fixing the BBH population to a fiducial set of parameters that correspond to the ``{\textsc{Power Law + Peak}}'' model (see Sec. \ref{sec:massDist}) and determining the values of $d_L$ (given $d_L^{\rm{GW}}$) as a function of $D$ while fixing $R_c$ and $n$ constant, we show how the detectable fraction of mergers depends on the number of spacetime dimensions in the top panel of Figure \ref{Fig:pdetz_dims}. At fixed $R_c=100 \rm \ Mpc$ and $R_c=1000 \rm \ Mpc$ with $n=2$, and we note that as $D$ increases from its GR value ($D=4$), the maximum detectable redshift as well as its peak value decreases for the detectable population. This makes sense, as energy from GW sources leaks for higher dimensional theories, leading to a reduction in their measured amplitude and subsequently a lower detectable horizon.

Similarly, we show the dependence of detectability as a function of $R_c$ in the bottom panel of Figure \ref{Fig:pdetz_dims}, while we fix the spacetime dimensions to $D=5$ and consider $n=2$ and $n=50$ as examples for the transition steepness. As $R_c$ increases, the maximum and peak redshift for the detectable distribution increases, where for $R_c \approx 10^4 \rm \ Mpc$ the model resembles GR-like behavior for $D=5$ and is independent of the value for $n$. If there is gravitational leakage, we expect the size of $R_c$ to be of cosmological scales, since small screening scales ($R_c<20 \rm / Mpc$) have been ruled out by the analysis of GW170817 in \cite{Pardo:2018ipy} irrespective of the value for $n$.

Finally, it is also worth investigating theories in which the graviton has a finite lifetime, meaning that it decays away as it travels cosmological distances. In this case, the modified luminosity distance of the GW would scale as \citep{Pardo:2018ipy}:
\begin{equation}
\label{eqn:decay}
d_L^{\rm{GW}} = d_L e^{d_L/R_g}
\end{equation}
where $R_g$ is the ``decay-length". Consequently, one can measure the graviton's ``decay-time"' as long as we assume that they propagate at the speed of light, that is, $t_g = R_g/c$. We leave constraints from decaying graviton models as future work.

\section{Methods}

\subsection{Hierarchical Inference} \label{sec:heirarchical inference}
We use hierarchical Bayesian inference to simultaneously infer the parameters for the population distribution of binary black hole mergers as well as the additional parameters that describe the gravitational leakage models described in Section \ref{sec:leak}. The binary black hole observations provide us with an estimate of their primary mass $m_1^{\text{det}}$ and mass ratio $q$ in the detector frame, as well as their luminosity distances $d_L$. 

The number density of BBH events as a function of detector frame quantities $(m_1^{\text{det}},m_2^{\text{det}},d_L)$ is related to the source frame parameters $(m_1,q,z)$ by,

\begin{equation} \label{number_dist}
    \frac{dN(m_1^{\text{det}},q,d_L | \Lambda)}{dm_1^{\text{det}} m_2^{\text{det}} dd_L} = \frac{1}{m_1(1+z)^2}\frac{dz}{dd_L}  \frac{dN(m_1, q, z | \Lambda)}{dm_1 dq dz}
\end{equation}
\noindent
with $\Lambda$ being the population hyperparameters that we want to measure and the proportionality factor relating detector frame to source frame quantities is the Jacobian transformation relating these parameterizations.

Now we relate the BBH number density in terms of the BBH merger rate density:
\begin{equation}
     \frac{dN(m_1, q, z | \Lambda)}{dm_1dqdz} = \frac{dV_c}{dz}\bigg(\frac{T_\mathrm{obs}}{1+z}\bigg) \frac{d\mathcal{R}(m_1, q, z | \Lambda)}{dm_1dq}
\end{equation}
\noindent
where the BBH merger rate $d\mathcal{R}$ over a range of primary mass, mass ratio and redshift (assuming the BBH mass distribution is redshift independent) gives:
\begin{equation} \label{number_density}
    \frac{d\mathcal{R}(m_1, q, z | \Lambda)}{dm_1dq} = \mathcal{R}_0 p(m_1, q | \Lambda)p(z | \Lambda),
\end{equation}
\noindent
here $\mathcal{R}_0$ is the local merger rate at $z=0$. The BBH population is modeled through the normalized mass distribution $p(m_1,q|\Lambda) = p(m_1|\Lambda)p(q|m_1,\Lambda)$ and its redshift evolution $p(z|\Lambda)$, which is chosen such that $p(z=0|\Lambda)=1$. Here $dV_c/dz$ is the differential uniform-in-comoving volume element, $T_\mathrm{obs}$ the total observation time and the factor of $1/(1+z)$ converts source-frame time to detector-frame time. By integrating the BBH number density across all primary masses and mass ratios, and out to a maximum redshift $z_\mathrm{max}$ we get the expected number of BBH within $z_\mathrm{max}$. In this work we take $z_\mathrm{max} = 3$ throughout. 

Given a set of $N_\mathrm{obs}$ gravitational wave observations $\{d_i\}$, we can calculate the posterior on $\Lambda$ following e.g.~\citet{Farr_2019} and \citet{Mandel_2019}:
\begin{align} \label{eqn:posterior}
\begin{split}
    p\left(\Lambda | \{d_i\} \right)  &\propto p(\Lambda) e^{-\mathcal{R}_0 \xi(\Lambda)} \prod_{i=1}^{N_\mathrm{obs}} \Bigg[ \int \mathcal{L}\left(d_i | m_1^i, q^i, z^i \right) \\
    & \times \frac{dN(m_1, q, z | \Lambda)}{dm_1dqdz} dm_1 dq dz \Bigg],
\end{split}
\end{align}
\noindent
where $\mathcal{L}(d_i|m_1, q, z)$ is the single-event likelihood function for each event, and $\xi(\Lambda)$ is the detectable fraction of sources corresponding to a population determined by the population hyperparameters $\Lambda$. 

Following \cite{Farr_2019} to estimate $\xi(\Lambda)$, we assume that sampling of $\xi$ will follow a normal distribution (i.e. $\xi(\Lambda) \sim \mathcal{N}(\Lambda|\mu, \sigma)$), with $\mu$ the importance sample estimate of $\xi$ and $\sigma$ its associated uncertainty.
Practically, we estimate the detectable fraction $\xi(\Lambda)$, by using the LVK's injection campaign of BBH events simulated from a broad BBH population and injected into real detector data, then searched for using the same analysis pipelines that found GWTC-3 and prior GW Transient catalogs \citep{LIGOScientific:2021psn}. 

Finally, we marginalize over the local BBH merger rate $\mathcal{R}_0$ using a log-Uniform prior \citep{o3a_pop}, and neglect terms of $\mathcal{O}(N_\mathrm{eff}^{-2})$ \citep{Farr_2019}. We also approximate the integral over the individual event likelihoods in Equation \ref{eqn:posterior} with importance sampling over $N_i$ single-event posterior samples generated from single event inference analysis with default event prior $\pi(m_1, q, z) \propto d_L^2 m_1 (1+z)^2 \frac{dd_L}{dz}.$ \footnote{Note that we have absorbed the Jacobian transformation in Equation \ref{number_dist} in the definition for $\pi(m_1, q, z)$ as it is typically done.}

\subsection{Binary Black Hole Population Models}
\label{sec:massDist}
We model the black hole mass distribution in the source frame using the ``{\textsc{Power Law + Peak}}'' model as described the LVK second gravitational-wave transient catalog (GWTC-2) population analysis \citep{o3a_pop,o3pop} and in \cite{Talbot_2018}. It is a mixture distribution consisting of a power law distribution truncated at a maximum mass to model the PISNe mass gap and of a Gaussian distribution that models the build-up of black holes due to the PPISNe (pulsational PISNe) mass loss \citep{Talbot_2018}. For simplicity, we neglect the low-mass smoothing feature used in \citep{o3a_pop}. 

Under this model, the probability distribution for the primary mass is,
\begin{align}
\begin{split}\label{pm1}
p(m_1 | \Lambda) &= (1-f_{\rm{pp}})p(m_1|\alpha, m_{\rm min}, m_{\rm{max}}) \\
                       & + f_{\rm{pp}} p_{\rm{pp}}(m_1|m_{\rm{pp}},\sigma_{\rm{pp}})\, ,
\end{split}
\end{align}
where $f_{\rm pp}$ is a mixing fraction parameter that gives the weight of the Gaussian component \citep{o3a_pop}.

The power law distribution is defined as,
\begin{equation}
p(m_1|\alpha, m_{\rm{max}})  \propto {({m_1})}^{-\alpha} \, ,
\end{equation}
where $\alpha$ is the powerlaw index and $m_{\rm{min}}$ ($m_{\rm{max}}$) is the minimum (maximum) black-hole mass with the constraint that $m_{\rm{min}} < m_1 < m_{\rm{max}}$. The Gaussian component has a mean $m_{\rm{pp}}$ and a standard deviation $\sigma_{\rm{pp}}$ and is given by,
\begin{equation}
p(m_1|m_{\rm{pp}},\sigma_{\rm{pp}}) \propto \exp\left[-\frac{({m_1} - m_{\rm{pp}})^2}{2\sigma_{\rm{pp}}^2}\right] \, .
\end{equation}

Following \cite{o3a_pop}, we assume a powerlaw distribution for the mass ratio ($q = m_{2}/m_{1}\leq 1$) (with powerlaw index $\beta$). So the conditional probability distribution on $q$ given $m_1$ can be written as,
\begin{equation}\label{pm2}
p(q|m_1, \beta, m_{\rm{min}} )\propto q^{\beta}. 
\end{equation}
and is defined in the range $m_{\rm min}/m_1 < q < 1$.

The redshift distribution for the binary black hole population is assumed to be a distribution that closely resembles the star formation rate but it is flexible enough to accommodate different shapes. Hence we follow the model used in \cite{Callister:2020arv,Ezquiaga:2021ayr},
\begin{equation}\label{pm2}
p(z|\gamma, \kappa, z_p) \propto \frac{(1+z)^\gamma}{1+\left(\frac{1+z}{1+z_p}\right)^{\gamma+\kappa}}
\end{equation}
which peaks at $z_p$ and where $\gamma$ controls the low redshift rise and $\kappa$ the high redshift tail of the distribution.

\section{Results}
 We present constraints on gravitational leakage models using GWTC-3 BBH mergers. We use only the events that pass the 1 per-year IFAR threshold as was done in \cite{LIGOScientific:2021psn}. In total we analyze 69 BBH mergers, and exclude outlier BBH events such as GW190814 or any NSBH and BNS candidates.
 
 \begin{figure*}[htb]
\includegraphics[width=0.48\textwidth]{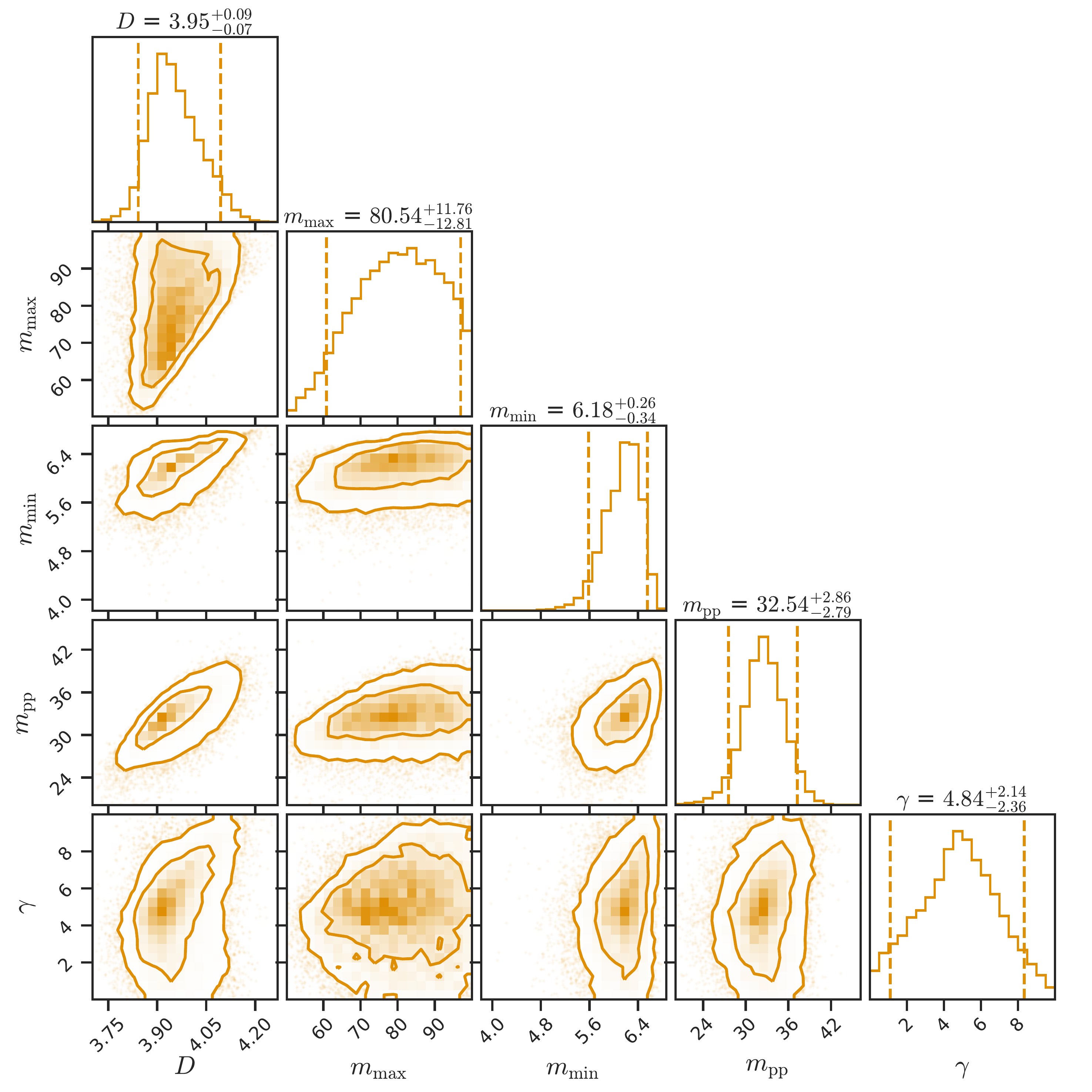}
\includegraphics[width=0.48\textwidth]{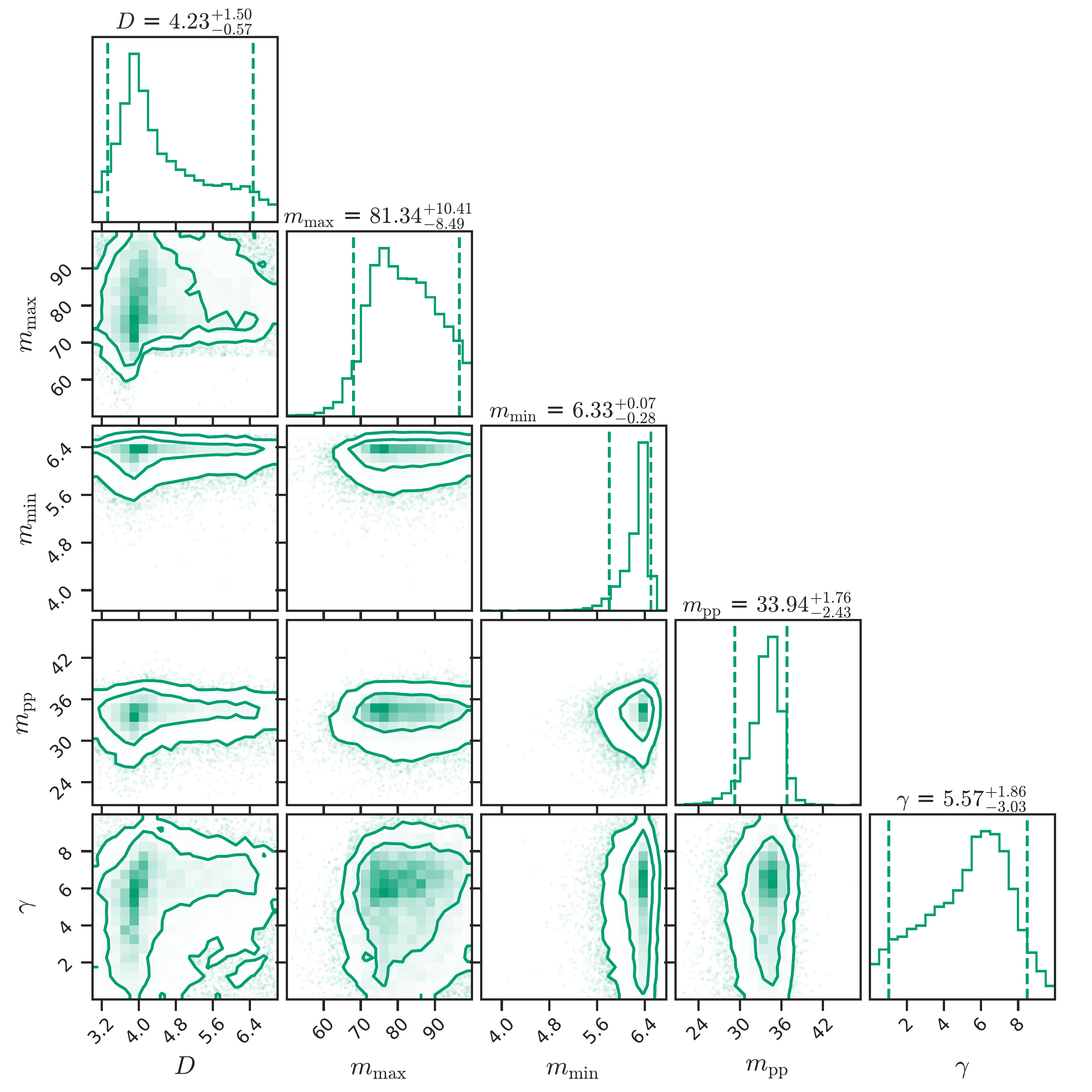}
\caption{Posterior distributions for the number of spacetime dimensions $D$ and the BBH population parameters most strongly correlated with $D$: the BBH maximum mass $m_{\rm max}$, minimum mass $m_{\rm max}$, peak of the Gaussian (PISN) component $m_{\rm pp}$ and the slope of the merger rate evolution $\gamma$. Left panel: We show the posterior for the gravitational leakage model defined in Equation \ref{eqn:grwaveform}. Right panel: Posterior distribution for the leakage model with screening scale $R_c$ and transition steepness $n$ as defined in \ref{eqn:defmenwaveform}.}
\label{Fig:posterior_small}
\end{figure*}

First we show constraints for the gravitational leakage model that only depends on the number of spacetime dimensions $D$ (see Equation \ref{eqn:grwaveform}) and so allows for gravitational leakage at all scales. The posterior distribution on the number of spacetime dimensions $D$ as well as the hyperparameters that correlate strongly with $D$ are shown in the left-panel of Figure \ref{Fig:posterior_small} (See Appendix \ref{sec:posteriors} for full posterior distributions). We find that we can place tight constraints on the number of spacetime dimensions with BBH observations alone to $D=3.95^{+0.09}_{-0.07}$ at $68\%$ C.L. The constraint presented here is as competitive to the GW170817 constraint with its associated electromagnetic counterpart of $D=3.98^{+0.07}_{-0.09}$ at $68\%$ C.L. \citep{Pardo:2018ipy}. The BBH population hyperparameters are broadly consistent with constraints placed in the GWTC-3 population analysis by the LVK Collaboration. However, we do find a broadening and shift to higher mass for the allowed BBH maximum mass $m_{\rm max}$. This is consistent with the analysis of \citep{Ezquiaga:2020tns}, as the additional  parameters in the modified luminosity distance models due to gravitational leakage (or varying $c_M$) introduce uncertainty in the value of $m_{\rm max}$.

We also show results for the model of Equation \ref{eqn:defmenwaveform} that allows for a varying screening scale and transition steepness. Posterior distributions on $D$ and relevant population parameters are shown in the right hand side panel of Figure \ref{Fig:posterior_small}. Under this model, we find we find $D=4.23^{+1.50}_{-0.57}$ at $68\%$ C.L., which gives broader constraints compared to the previous model (due to the additional screening scale parameters) but are still consistent with GR. We find that $D$ has strong correlations with the screening scale and transition steepness as shown in Figure \ref{Fig:postB} which is also explained through the detectability of the simulated population $p(\textrm{det}|z)$ shown in Figure \ref{Fig:pdetz_dims}. We note that, the BBH population hyperparameters under this model are similarly constrained with respect to the model of Equation \ref{eqn:grwaveform}.

\begin{figure}[htb]
\includegraphics[width=0.5\textwidth]{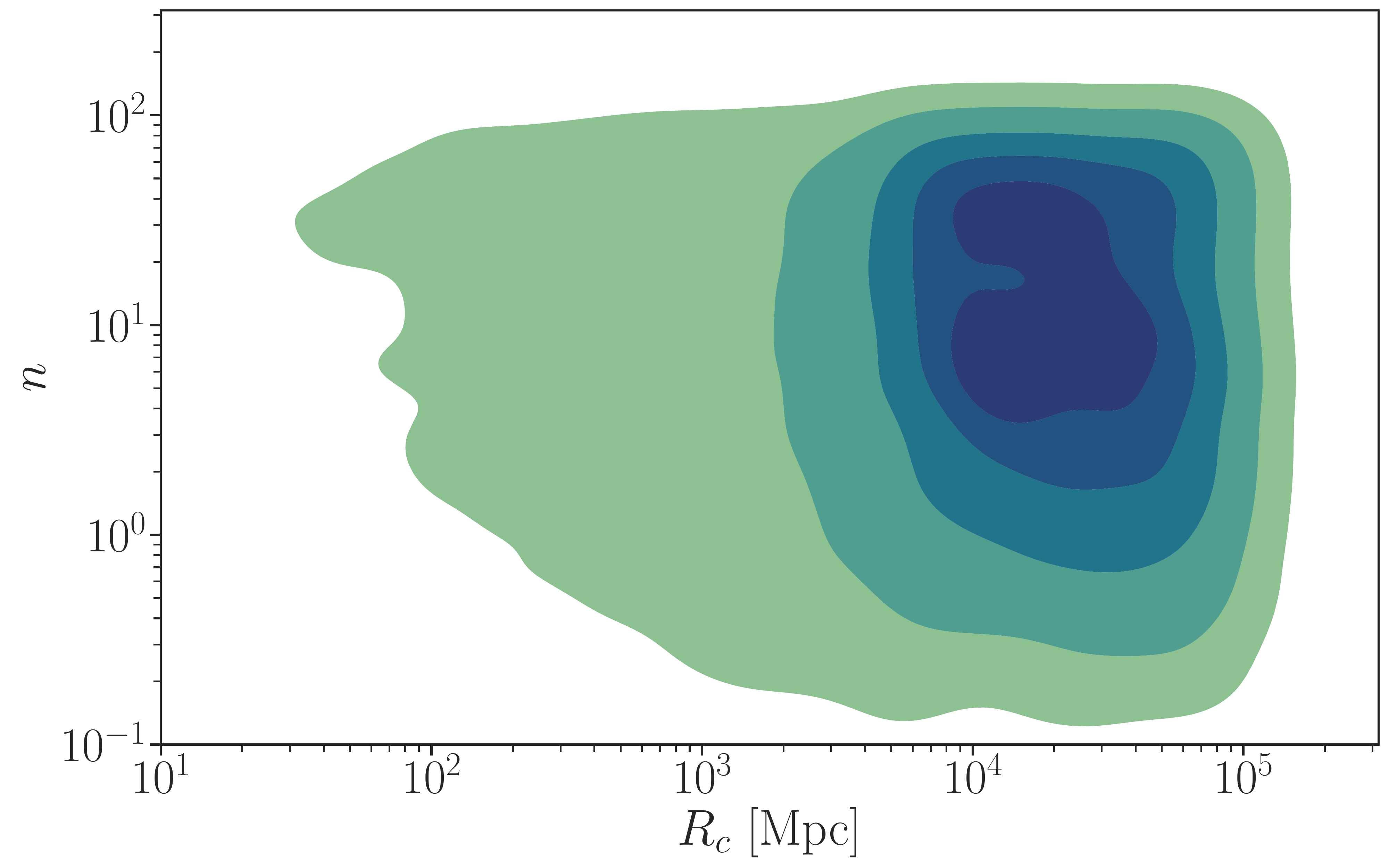}
\caption{Constraints on the screening scale and transition steepness from GWTC-3 BBH observations. We find that $\log_{10} R_c/{\rm Mpc}= 4.14^{+0.55}_{-0.86}$  with a transition steepness $\log_{10} n = 0.86^{+0.73}_{-0.84}$ C.L., consistent with previous upper limits placed by GW170817 and GW190521 \citep{Pardo:2018ipy,Corman:2021avn}. We find that these parameters are consistent with GR, e.g. large screening scale.}
\label{Fig:screening}
\end{figure}

Since we have used binary black hole mergers across a wide range of distances, we are also able to constrain the screening scale $R_c$ and the transition steepness $n$ for the first time, without having to fix any of the three model parameters as has been done in other studies (See Appendix \ref{sec:posteriors} for full posterior distributions). We show in Figure \ref{Fig:screening}, the posterior distribution on $R_c$ and $n$ and find that $\log_{10} R_c/{\rm Mpc}= 4.14^{+0.55}_{-0.86}$ with a transition steepness $\log_{10} n = 0.86^{+0.73}_{-0.84}$ both ath $68\%$ C.L., consistent with GW170817 (and GW190521) constraints \citep{Pardo:2018ipy,Corman:2021avn}. These constraints are also consistent with General Relativity since in the limit of large $R_c$ we expect GR-like behavior independently of $n$.

We also report Bayes factors $\ln \mathcal{B}^{D}_{\rm GR}$, to do hypothesis testing while fixing the number of spacetime dimensions $D = 4$ (GR) compared to $D \neq 4$, under the leakage model of Equation \ref{eqn:defmenwaveform}. We report the following Bayes factors, $\mathcal{B}^{D=5}_{\rm GR} = -0.47$, $\mathcal{B}^{D=6}_{\rm GR} = -1.18$, $\mathcal{B}^{D=7}_{\rm GR} = -1.21$ and $\mathcal{B}^{D=8}_{\rm GR} = -1.39$. As expected, higher number of spacetime dimensions with respect to $D=4$ are disfavored by the data.

\section{Discussion}
We placed constraints on the number of spacetime dimensions from binary black hole mergers using the recently released GWTC-3 catalog. We find that our constraints on $D$ agree with general relativity for both gravitational leakage models considered in this work. We also find that the constraints placed on the screening scale and transition steepness for gravitational leakage models are consistent with General Relativity. We note that, further, and more distant observations should allow us to place tighter constraints on these scales. 

The constraints presented in this paper are dependent on the chosen binary black hole mass and redshift distribution models. However, the phenomenological models used in this work should be flexible enough to capture the astrophysics that drives different BBH formation channels at current detector sensitivities. The BBH population modeling can be improved to include sub-populations such as hierarchical mergers or even mergers above the PISNe mass gap, we leave this study to future work. 

With respect to gravity leakage models, one can use the Bayesian framework presented here to test different parameterizations for the modified luminosity distance $d_L^{\rm GW}$ under different models. For example, alternative phenomenological models such as the graviton leakage model of \cite{Pardo:2018ipy} or theory specific models from specific parameterizations of modified gravity theories.

The Bayesian analysis used in this work, can also be extended to work with the dark siren formalism that has been used mainly to perform cosmological measurements on $H_0$, since it provides a discrete prior on the allowed locations (assuming they are hosted by a galaxy) for GW sources.

Binary black hole mergers can probe cosmology as well as extensions of the $\Lambda$CDM cosmological model under modified gravity at cosmological scales without the need of an independent redshift measurement to the source or other electromagnetic information. With future, more distant GW catalogs and an improved understanding of the BBH population we can only expect to improve these measurements and better understand the underlying cosmological model.

\section*{Acknowledgements} 
The author would like to thank Amanda Baylor, Patrick Brady, Jolien Creighton, Rico Ka Lok Lo, Soichiro Morisaki and Daniel Wysocki for useful comments and feedback throughout this work. IMH is supported by the NSF Graduate Research Fellowship Program under grant DGE-17247915. This work was supported by NSF awards PHY-1912649. The author is grateful for computational resources provided by the Leonard E Parker Center for Gravitation, Cosmology and Astrophysics at the University of Wisconsin-Milwaukee. We thank LIGO and Virgo Collaboration for providing the data for this work. This research has made use of data, software and/or web tools obtained from the Gravitational Wave Open Science Center (https://www.gw-openscience.org/ ), a service of LIGO Laboratory, the LIGO Scientific Collaboration and the Virgo Collaboration. LIGO Laboratory and Advanced LIGO are funded by the United States National Science Foundation (NSF) as well as the Science and Technology Facilities Council (STFC) of the United Kingdom, the Max-Planck-Society (MPS), and the State of Niedersachsen/Germany for support of the construction of Advanced LIGO and construction and operation of the GEO600 detector. Additional support for Advanced LIGO was provided by the Australian Research Council. Virgo is funded, through the European Gravitational Observatory (EGO), by the French Centre National de Recherche Scientifique (CNRS), the Italian Istituto Nazionale di Fisica Nucleare (INFN) and the Dutch Nikhef, with contributions by institutions from Belgium, Germany, Greece, Hungary, Ireland, Japan, Monaco, Poland, Portugal, Spain. This material is based upon work supported by NSF's LIGO Laboratory which is a major facility fully funded by the National Science Foundation. This article has been assigned LIGO document number LIGO-P2100449.

\appendix
\section{Full Posterior Distributions}
\label{sec:posteriors}

\begin{figure}[htb]
\includegraphics[width=\textwidth]{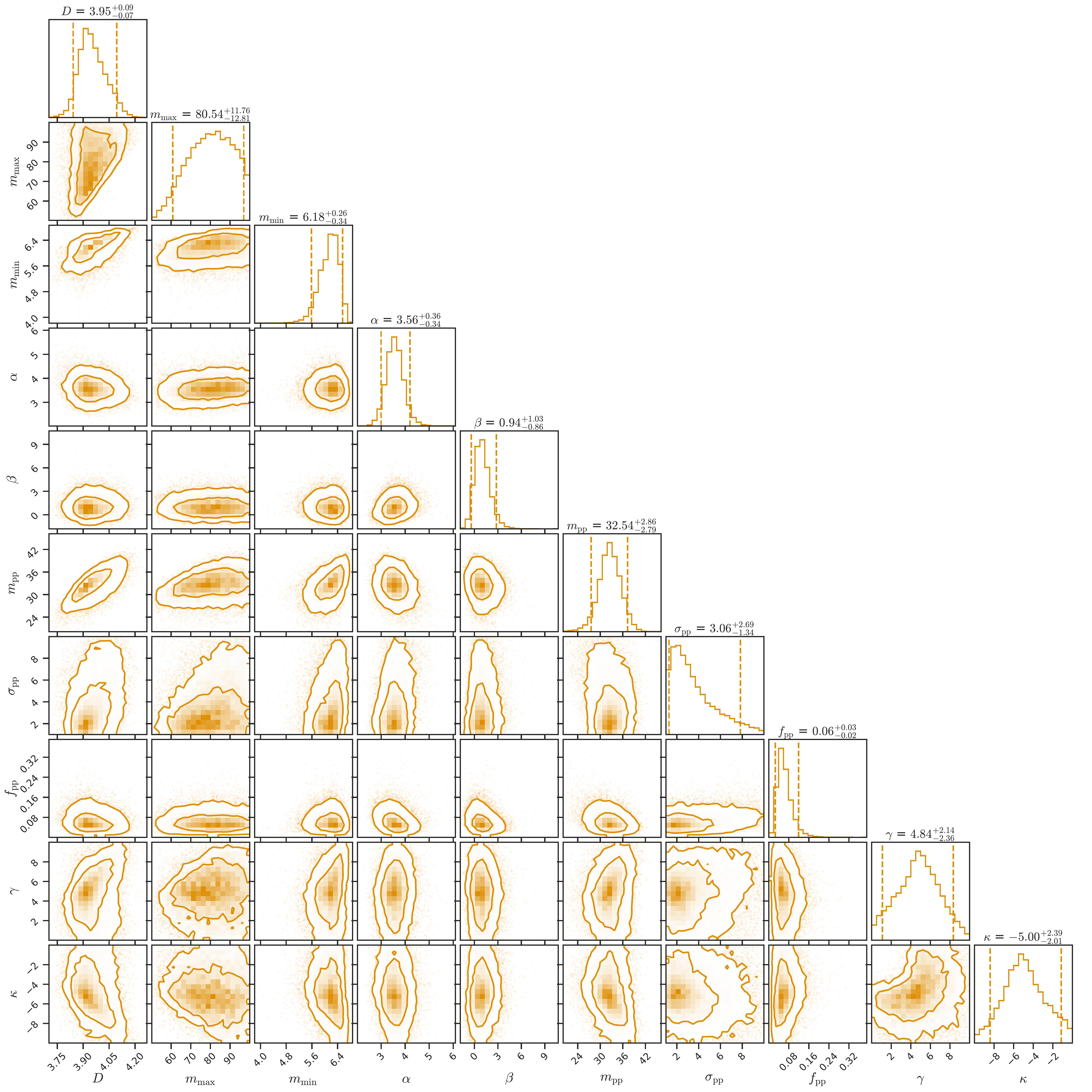}
\caption{Posterior distribution on number of spacetime dimensions $D$ using the model in Equation \ref{eqn:grwaveform}, as well as the ``{\textsc{Power Law + Peak}}'' mass distribution parameters $m_{\rm min}$, $m_{\rm max}$, $\alpha$, $\beta$, $m_{\rm pp}$, $\sigma_{\rm pp}$ and $f_{\rm pp}$ and the SFR-like redshift evolution model with parameters $\gamma$, $\kappa$ and $z_{\rm peak}$. }
\label{Fig:postA}
\end{figure}

\begin{figure}[htb]
\includegraphics[width=\textwidth]{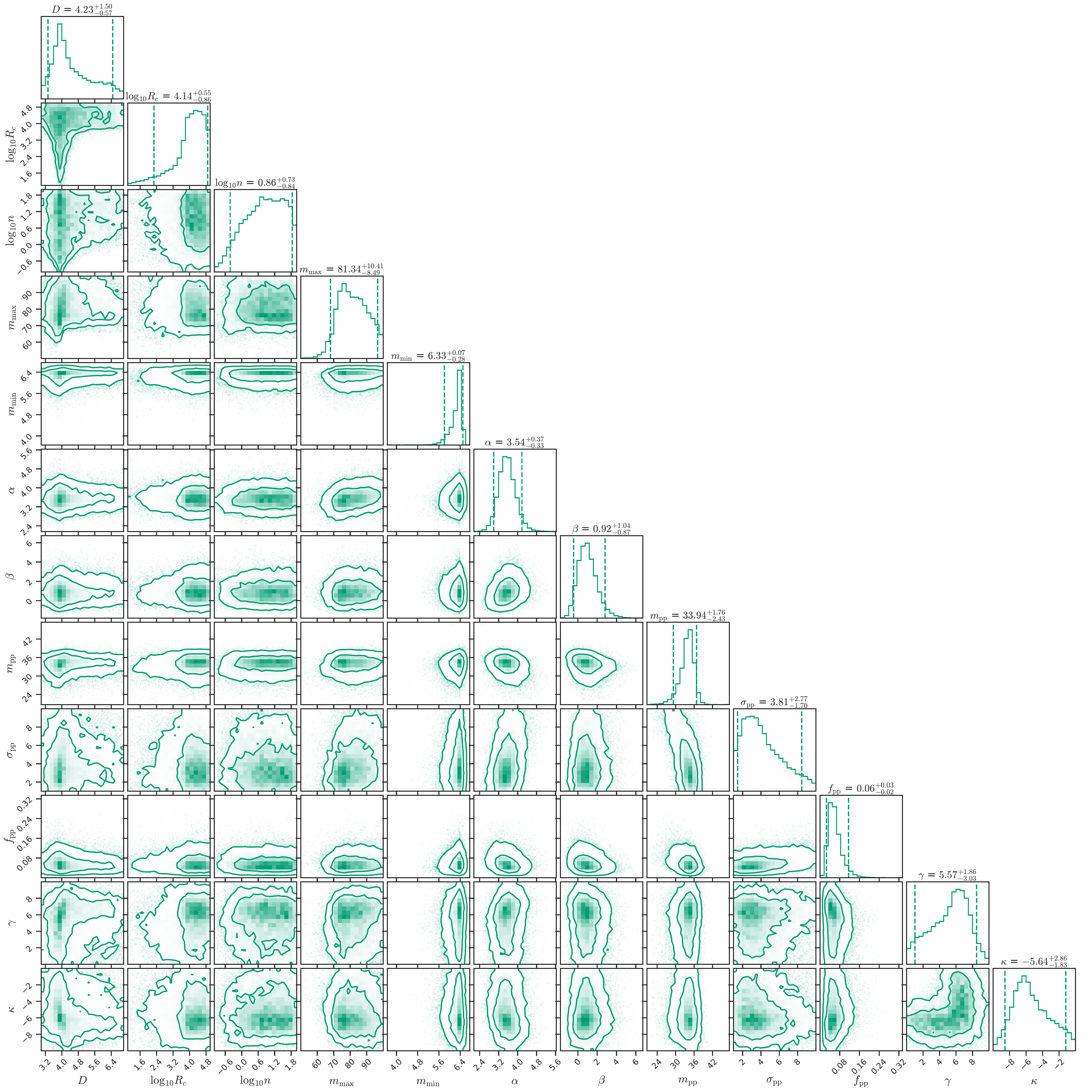}
\caption{Posterior distribution on number of spacetime dimensions $D$, screening scale $R_c$ and transition steepness $n$ using the model in Equation \ref{eqn:defmenwaveform}, as well as the ``{\textsc{Power Law + Peak}}'' mass distribution parameters the SFR-like redshift evolution model with parameters. }
\label{Fig:postB}
\end{figure}

\begin{thebibliography}{}
\expandafter\ifx\csname natexlab\endcsname\relax\def\natexlab#1{#1}\fi
\providecommand{\url}[1]{\href{#1}{#1}}
\providecommand{\dodoi}[1]{doi:~\href{http://doi.org/#1}{\nolinkurl{#1}}}
\providecommand{\doeprint}[1]{\href{http://ascl.net/#1}{\nolinkurl{http://ascl.net/#1}}}
\providecommand{\doarXiv}[1]{\href{https://arxiv.org/abs/#1}{\nolinkurl{https://arxiv.org/abs/#1}}}

\bibitem[{Aasi {et~al.}(2015)}]{LIGOScientific:2014pky}
Aasi, J., {et~al.} 2015, Class. Quant. Grav., 32, 074001,
  \dodoi{10.1088/0264-9381/32/7/074001}

\bibitem[{Abbott {et~al.}(2017{\natexlab{a}})}]{TheLIGOScientific:2017qsa}
Abbott, B., {et~al.} 2017{\natexlab{a}}, Phys. Rev. Lett., 119, 161101,
  \dodoi{10.1103/PhysRevLett.119.161101}

\bibitem[{Abbott {et~al.}(2019{\natexlab{a}})}]{Abbott:2019yzh}
---. 2019{\natexlab{a}}.
\newblock \doarXiv{1908.06060}

\bibitem[{Abbott {et~al.}(2021{\natexlab{a}})}]{LIGOScientific:2021psn}
---. 2021{\natexlab{a}}.
\newblock \doarXiv{2111.03634}

\bibitem[{Abbott {et~al.}(2017{\natexlab{b}})}]{Abbott:2017xzu}
Abbott, B.~P., {et~al.} 2017{\natexlab{b}}, Nature, 551, 85,
  \dodoi{10.1038/nature24471}

\bibitem[{Abbott {et~al.}(2019{\natexlab{b}})}]{gwtc1}
---. 2019{\natexlab{b}}, Phys. Rev. X, 9, 031040,
  \dodoi{10.1103/PhysRevX.9.031040}

\bibitem[{Abbott {et~al.}(2021{\natexlab{b}})}]{o2cosmo}
---. 2021{\natexlab{b}}, Astrophys. J., 909, 218,
  \dodoi{10.3847/1538-4357/abdcb7}

\bibitem[{Abbott {et~al.}(2020)Abbott, Abbott, Abraham, Acernese, Ackley,
  Adams, Adams, Adhikari, Adya, Affeldt, \& et~al.}]{o3a_pop}
Abbott, R., Abbott, T.~D., Abraham, S., {et~al.} 2020, Population Properties of
  Compact Objects from the Second LIGO-Virgo Gravitational-Wave Transient
  Catalog.
\newblock \doarXiv{2010.14533}

\bibitem[{Abbott {et~al.}(2021{\natexlab{c}})}]{gwtc2}
Abbott, R., {et~al.} 2021{\natexlab{c}}, Phys. Rev. X, 11, 021053,
  \dodoi{10.1103/PhysRevX.11.021053}

\bibitem[{Abbott {et~al.}(2021{\natexlab{d}})}]{gwtc21}
---. 2021{\natexlab{d}}.
\newblock \doarXiv{2108.01045}

\bibitem[{Abbott {et~al.}(2021{\natexlab{e}})}]{gwtc3}
---. 2021{\natexlab{e}}.
\newblock \doarXiv{2111.03606}

\bibitem[{Abbott {et~al.}(2021{\natexlab{f}})}]{o3pop}
---. 2021{\natexlab{f}}.
\newblock \doarXiv{2111.03634}

\bibitem[{Abbott {et~al.}(2021{\natexlab{g}})}]{o3cosmo}
---. 2021{\natexlab{g}}.
\newblock \doarXiv{2111.03604}

\bibitem[{Abbott {et~al.}(2021{\natexlab{h}})}]{o3lens}
---. 2021{\natexlab{h}}.
\newblock \doarXiv{2105.06384}

\bibitem[{Abbott {et~al.}(2021{\natexlab{i}})}]{o3atgr}
---. 2021{\natexlab{i}}, Phys. Rev. D, 103, 122002,
  \dodoi{10.1103/PhysRevD.103.122002}

\bibitem[{Abbott {et~al.}(2021{\natexlab{j}})}]{LIGOScientific:2021sio}
---. 2021{\natexlab{j}}.
\newblock \doarXiv{2112.06861}

\bibitem[{Acernese {et~al.}(2015)}]{VIRGO:2014yos}
Acernese, F., {et~al.} 2015, Class. Quant. Grav., 32, 024001,
  \dodoi{10.1088/0264-9381/32/2/024001}

\bibitem[{Callister {et~al.}(2020)Callister, Fishbach, Holz, \&
  Farr}]{Callister:2020arv}
Callister, T., Fishbach, M., Holz, D., \& Farr, W. 2020, Astrophys. J. Lett.,
  896, L32, \dodoi{10.3847/2041-8213/ab9743}

\bibitem[{Chen {et~al.}(2018)Chen, Fishbach, \& Holz}]{Chen:2017rfc}
Chen, H.-Y., Fishbach, M., \& Holz, D.~E. 2018, Nature, 562, 545,
  \dodoi{10.1038/s41586-018-0606-0}

\bibitem[{Corman {et~al.}(2021)Corman, Ghosh, Escamilla-Rivera, Hendry, Marsat,
  \& Tamanini}]{Corman:2021avn}
Corman, M., Ghosh, A., Escamilla-Rivera, C., {et~al.} 2021.
\newblock \doarXiv{2109.08748}

\bibitem[{Deffayet \& Menou(2007)}]{Deffayet:2007kf}
Deffayet, C., \& Menou, K. 2007, Astrophys. J. Lett., 668, L143,
  \dodoi{10.1086/522931}

\bibitem[{Del~Pozzo(2012)}]{DelPozzo:2011yh}
Del~Pozzo, W. 2012, Phys. Rev., D86, 043011, \dodoi{10.1103/PhysRevD.86.043011}

\bibitem[{Diaz \& Mukherjee(2021)}]{Diaz:2021pem}
Diaz, C.~C., \& Mukherjee, S. 2021.
\newblock \doarXiv{2107.12787}

\bibitem[{Ezquiaga(2021)}]{Ezquiaga:2021ayr}
Ezquiaga, J.~M. 2021, Phys. Lett. B, 822, 136665,
  \dodoi{10.1016/j.physletb.2021.136665}

\bibitem[{Ezquiaga \& Holz(2021)}]{Ezquiaga:2020tns}
Ezquiaga, J.~M., \& Holz, D.~E. 2021, Astrophys. J. Lett., 909, L23,
  \dodoi{10.3847/2041-8213/abe638}

\bibitem[{Farr(2019)}]{Farr_2019}
Farr, W.~M. 2019, Research Notes of the AAS, 3, 66,
  \dodoi{10.3847/2515-5172/ab1d5f}

\bibitem[{Finke {et~al.}(2021)Finke, Foffa, Iacovelli, Maggiore, \&
  Mancarella}]{Finke:2021aom}
Finke, A., Foffa, S., Iacovelli, F., Maggiore, M., \& Mancarella, M. 2021,
  JCAP, 08, 026, \dodoi{10.1088/1475-7516/2021/08/026}

\bibitem[{Fishbach {et~al.}(2019)Fishbach, Gray, Magaña~Hernandez, Qi, \&
  Sur}]{Fishbach:2018gjp}
Fishbach, M., Gray, R., Magaña~Hernandez, I., Qi, H., \& Sur, A. 2019,
  Astrophys. J., 871, L13, \dodoi{10.3847/2041-8213/aaf96e}

\bibitem[{Gray {et~al.}(2019)Gray, Magaña~Hernandez, Qi, Sur,
  {et~al.}}]{Gray:2019ksv}
Gray, R., Magaña~Hernandez, I., Qi, H., Sur, A., {et~al.} 2019.
\newblock \doarXiv{1908.06050}

\bibitem[{Heger {et~al.}(2003)Heger, Fryer, Woosley, Langer, \&
  Hartmann}]{Heger_2003}
Heger, A., Fryer, C.~L., Woosley, S.~E., Langer, N., \& Hartmann, D.~H. 2003,
  The Astrophysical Journal, 591, 288–300, \dodoi{10.1086/375341}

\bibitem[{Heger \& Woosley(2002)}]{Heger_2002}
Heger, A., \& Woosley, S.~E. 2002, The Astrophysical Journal, 567, 532–543,
  \dodoi{10.1086/338487}

\bibitem[{Holz \& Hughes(2005)}]{Holz:2005df}
Holz, D.~E., \& Hughes, S.~A. 2005, Astrophys. J., 629, 15,
  \dodoi{10.1086/431341}

\bibitem[{Lagos {et~al.}(2019)Lagos, Fishbach, Landry, \&
  Holz}]{PhysRevD.99.083504}
Lagos, M., Fishbach, M., Landry, P., \& Holz, D.~E. 2019, Phys. Rev. D, 99,
  083504, \dodoi{10.1103/PhysRevD.99.083504}

\bibitem[{Mancarella {et~al.}(2021)Mancarella, Genoud-Prachex, \&
  Maggiore}]{Mancarella:2021ecn}
Mancarella, M., Genoud-Prachex, E., \& Maggiore, M. 2021.
\newblock \doarXiv{2112.05728}

\bibitem[{Mandel {et~al.}(2019)Mandel, Farr, \& Gair}]{Mandel_2019}
Mandel, I., Farr, W.~M., \& Gair, J.~R. 2019, Monthly Notices of the Royal
  Astronomical Society, 486, 1086–1093, \dodoi{10.1093/mnras/stz896}

\bibitem[{Mukherjee {et~al.}(2021{\natexlab{a}})Mukherjee, Wandelt, Nissanke,
  \& Silvestri}]{Mukherjee:2020hyn}
Mukherjee, S., Wandelt, B.~D., Nissanke, S.~M., \& Silvestri, A.
  2021{\natexlab{a}}, Phys. Rev. D, 103, 043520,
  \dodoi{10.1103/PhysRevD.103.043520}

\bibitem[{Mukherjee {et~al.}(2021{\natexlab{b}})Mukherjee, Wandelt, \&
  Silk}]{Mukherjee:2020mha}
Mukherjee, S., Wandelt, B.~D., \& Silk, J. 2021{\natexlab{b}}, Mon. Not. Roy.
  Astron. Soc., 502, 1136, \dodoi{10.1093/mnras/stab001}

\bibitem[{Nair {et~al.}(2018)Nair, Bose, \& Saini}]{Nair:2018ign}
Nair, R., Bose, S., \& Saini, T.~D. 2018, Phys. Rev., D98, 023502,
  \dodoi{10.1103/PhysRevD.98.023502}

\bibitem[{Palmese {et~al.}(2021)Palmese, Bom, Mucesh, \&
  Hartley}]{Palmese:2021mjm}
Palmese, A., Bom, C.~R., Mucesh, S., \& Hartley, W.~G. 2021.
\newblock \doarXiv{2111.06445}

\bibitem[{Palmese {et~al.}(2020)}]{DES:2020nay}
Palmese, A., {et~al.} 2020, Astrophys. J. Lett., 900, L33,
  \dodoi{10.3847/2041-8213/abaeff}

\bibitem[{Pardo {et~al.}(2018)Pardo, Fishbach, Holz, \&
  Spergel}]{Pardo:2018ipy}
Pardo, K., Fishbach, M., Holz, D.~E., \& Spergel, D.~N. 2018, JCAP, 07, 048,
  \dodoi{10.1088/1475-7516/2018/07/048}

\bibitem[{Schutz(1986)}]{Schutz:1986gp}
Schutz, B.~F. 1986, Nature, 323, 310, \dodoi{10.1038/323310a0}

\bibitem[{Soares-Santos {et~al.}(2019)Soares-Santos, Palmese,
  {et~al.}}]{Soares-Santos:2019irc}
Soares-Santos, M., Palmese, A., {et~al.} 2019, Astrophys. J., 876, L7,
  \dodoi{10.3847/2041-8213/ab14f1}

\bibitem[{Talbot \& Thrane(2018)}]{Talbot_2018}
Talbot, C., \& Thrane, E. 2018, The Astrophysical Journal, 856, 173,
  \dodoi{10.3847/1538-4357/aab34c}

\bibitem[{Taylor {et~al.}(2012)Taylor, Gair, \& Mandel}]{PhysRevD.85.023535}
Taylor, S.~R., Gair, J.~R., \& Mandel, I. 2012, Phys. Rev. D, 85, 023535,
  \dodoi{10.1103/PhysRevD.85.023535}

\bibitem[{Woosley(2017)}]{Woosley_2017}
Woosley, S.~E. 2017, The Astrophysical Journal, 836, 244,
  \dodoi{10.3847/1538-4357/836/2/244}

\bibitem[{Woosley(2019)}]{Woosley_2019}
---. 2019, The Astrophysical Journal, 878, 49, \dodoi{10.3847/1538-4357/ab1b41}

\bibitem[{Woosley {et~al.}(2002)Woosley, Heger, \& Weaver}]{PISN_Woosley}
Woosley, S.~E., Heger, A., \& Weaver, T.~A. 2002, Rev. Mod. Phys., 74, 1015,
  \dodoi{10.1103/RevModPhys.74.1015}

\end{thebibliography}
\end{document}